\documentclass[aps,prc]{revtex4-2}
\usepackage{amsmath,amssymb}
\usepackage{graphicx}

\newcommand{\be}{\begin{equation}}
\newcommand{\ee}{\end{equation}}
\newcommand{\ba}{\begin{eqnarray}}
\newcommand{\ea}{\end{eqnarray}}
\newcommand{\ban}{\begin{eqnarray*}}
\newcommand{\ean}{\end{eqnarray*}}

\begin{document}

\title{Deuteron-Deuteron Correlation Function in Nucleus-Nucleus Collisions}

\author{Stanis\l aw Mr\' owczy\' nski$^{1,2}$\footnote{e-mail: stanislaw.mrowczynski@ncbj.gov.pl} 
and Patrycja S\l o\'n$^1$\footnote{e-mail: pati4922@gmail.com}}

\affiliation{$^1$Institute of Physics, Jan Kochanowski University, ul. Uniwersytecka 7, 25-406 Kielce, Poland 
\\
$^2$National Centre for Nuclear Research, ul. Pasteura 7, 02-093 Warsaw, Poland}

\date{August 19, 2021}

\begin{abstract}

A formula of the $D$-$D$ correlation function is derived. The deuterons are treated either as elementary particles or as neutron-proton bound states. In the first case the deuterons are directly emitted from a source and in the second case the deuteron formation is a final-state process simultaneous with a generation of $D$-$D$ correlation. The source radius of deuterons formed due to final-state interactions is bigger by the factor of $\sqrt{2}$ than that of directly emitted deuterons. To check how sizable is the effect we compute the $D$-$D$ correlation function taking into the Bose-Einstein statistics of deuterons, the $s$-wave scattering due to the strong interaction and Coulomb repulsion. The correlation function is shown to be sensitive to the source radius for sources which are sufficiently small with RMS radii smaller than 3.5 fm. Otherwise the correlation function is dominated by the Coulomb repulsion and weakly depends on the source radius. Measurements which can make use of our finding are discussed. 

\end{abstract}

\maketitle

\section{Introduction}
\label{sec-into}

Deuterons are copiously produced in nucleus-nucleus collisions over a broad range of collision energies. They come mainly from nucleons that make up the initial nuclei, some of them are simply fragments of colliding nuclei. However, there are also genuine production processes if a generation of nucleon-antinucleon pairs is energetically possible. For this reason a  production of deuterons and antideuterons is observed even in proton-proton collisions at sufficiently high energies \cite{Alper:1973my,Henning:1977mt,Acharya:2019rgc,Acharya:2020sfy}. 

If the excitation energy per nucleon of matter produced in nucleus-nucleus collisions much exceeds the nuclear binding energy, one expects that deuterons are formed due to final-state interactions of nucleons emitted from the particle source. This is the coalescence mechanism of deuteron production proposed long ago \cite{Butler:1963pp,Schwarzschild:1963zz}, see also \cite{Mrowczynski:1987oid,Bellini:2020cbj}. If the excitation energy of produced matter is of the order of the nuclear binding energy, deuterons are mostly directly emitted from the interaction zone or they come from decays of heavier fragments. The coalescence of emitted nucleons is also possible. Small excitation energies of produced matter occur in low-energy nucleus-nucleus collisions and in the projectile and target fragmentation domains of high-energy collisions. 

Recently it has been argued that deuterons and other light nuclei are directly emitted not only from weakly excited nuclear matter but also from the extremely excited fireballs created in nucleus-nucleus collisions at RHIC and LHC. Specifically, the thermodynamical model has been found to predict very well the yields not only of hadron species measured at RHIC and the LHC but also of light nuclei and hypernuclei \cite{Andronic:2010qu,Cleymans:2011pe,Andronic:2017pug}. Since the yields are computed at the fireball freeze-out temperature, the agreement with experimental data suggests that the nuclei are directly emitted from the fireball. The result is very puzzling as one would not expect that loosely bound nuclei exist in the hot and dense fireball environment. Proponents of the thermal model speculate \cite{Andronic:2017pug} that nuclei are produced as colorless droplets of quarks and gluons with quantum numbers that match those of the final-state nuclei. In any case it is argued that the coalescence is not responsible for production of light nuclei. 

Mechanisms of deuteron production are under debate and, as reviewed in \cite{Mrowczynski:2020ugu}, some methods to resolve the problem experimentally have been proposed. Correlation functions of two hadrons with small relative momenta are particularly useful here as the functions are known to carry information about a space-time structure of hadrons' sources, see the review \cite{Lisa:2005dd}. We have shown very recently \cite{Mrowczynski:2019yrr} that a hadron-deuteron correlation function can tell us whether deuterons are directly emitted from the fireball or are formed afterwards due to final-state interactions. In the former case the radius of the source of deuterons coincides with that of nucleons. In the latter one the deuteron source radius is bigger by the factor of $\sqrt{4/3}$ because of a space-time extension of the deuteron formation process. The difference can be inferred from precisely measured proton-proton and proton-deutron correlation functions. 

In this paper we derive a formula of the deuteron-deuteron correlation function. At the beginning, in Sec.~\ref{sec-D-elementary}, deuterons are treated as elementary particles directly emitted from a source. Further on, in Sec.~\ref{sec-D-bound}, we assume that deuterons, which are explicitly treated as neutron-proton bound states, are formed only after nucleons are emitted from the source. The deuteron formation is thus simultaneous with a generation of deuteron-deuteron correlation. The main result of our analysis is that, if deuterons are formed due final-state interactions, the source radius is bigger by the factor $\sqrt{2}$ than that of directly emitted deuterons. 

In Sec.~\ref{sec-corr-fun} we compute the  $D$-$D$ correlation function to check whether the enlargement of the source radius is a measurable effect. We take into account an indistinguishability of the two deuterons, their $s$-wave scattering due to the strong interaction and Coulomb repulsion. We first compute the correlation functions for three spin states $S=0,1,2$ of the deuteron pair and the final correlation function is found as the average over the spin states. It is shown that the  $D$-$D$ correlation function is sensitive to the source radius for rather small sources. For bigger ones the correlation function is dominated by the Coulomb repulsion and becomes almost independent of the source radius. 

We are not aware of published results on the $D$-$D$ correlation function in high-energy collisions. However, the function was successfully measured in nucleus-nucleus collisions at the energy of a few tens of MeV per nucleon \cite{Chitwood:1985zz,Pochodzalla:1987zz,Gourio:2000tn} and in proton-nucleus collisions at 500 MeV \cite{Cebra:1989xe}, see also the review \cite{Verde:2006dh} and references therein. In Sec.~\ref{sec-corr-fun} we briefly discuss our results in the context of existing experimental data and we consider measurements that could make use our findings. 

We summarize our study and draw conclusions in Sec.~\ref{sec-conclusions}. Throughout the paper we use natural units with $\hbar = c = 1$.

\section{Deuterons treated as elementary particles}
\label{sec-D-elementary}

The $D$-$D$ correlation function $\mathcal{R}({\bf p}_1 , {\bf p}_2) $ is defined as 
\be
\frac{dP_{DD}}{d^3p_1 \, d^3p_2} = \mathcal{R}({\bf p}_1 , {\bf p}_2) \, 
\frac{dP_D}{d^3p_1} \frac{dP_D}{d^3p_2} ,
\ee
where $\frac{dP_D}{d^3p}$ is the probability density to observe a final-state deuteron with momentum ${\bf p}$ and $\frac{dP_{DD}}{d^3p_1 \, d^3p_2}$ is the probability density to observe a final-state pair of deuterons with momenta ${\bf p}_1$ and ${\bf p}_2$.

If we are interested in the correlations of deuterons with the relative momentum ${\bf q}$, which is much smaller than the momenta of each deuteron, that is ${\bf q} \ll {\bf p}_1$ and ${\bf q} \ll {\bf p}_2$, and the deuterons are treated as elementary particles, then the correlation function is given by the well-know formula (see {\it e.g.} the review \cite{Lisa:2005dd}) 
\be
\label{fun-corr-D-D-ele}
\mathcal{R}({\bf q}) 
= \int d^3 r_1 \, d^3 r_2  \, D({\bf r}_1) \, D({\bf r}_2)  
|\psi({\bf r}_1, {\bf r}_2)|^2 ,
\ee
where $\psi({\bf r}_1, {\bf r}_2)$ is the wave function of two deuterons with relative momentum ${\bf q}$ and $D({\bf r}) $ is the probability distribution of emission points of the deuterons normalized to unity $\int d^3 r \, D({\bf r}) =1$. 

The formula (\ref{fun-corr-D-D-ele}) is written as for the instantaneous emission of two deuterons as the source function is time independent. However, the time duration of the emission process $\tau$ can be easily incorporated \cite{Koonin:1977fh}. Then, one shows that in case of the isotropic source function we use the formula (\ref{fun-corr-D-D-ele}) still holds but the source radius is effectively enlarged due to the finite $\tau$, as discussed in detail in \cite{Maj:2009ue}. 

The source function usually depends on the momenta of emitted particles, see {\it e.g.} \cite{Adam:2015vja,Acharya:2020dfb}. We do not show the dependence but it is understood that the formula (\ref{fun-corr-D-D-ele}) holds for a limited interval of momenta of deuterons. 

We consider the deuteron-deuteron correlations in the center-of-mass frame of the pair and we treat the formula (\ref{fun-corr-D-D-ele}) as nonrelativistic even though the deuterons can be relativistic in both the rest frame of the source and in the laboratory frame. However, the correlation function significantly differs from unity only for small relative momenta. Therefore, the relative motion can be treated as nonrelativistic and the corresponding wave function is a solution of the Schr\"odinger equation. The source function, which is usually defined in the source rest frame, needs to be transformed to the center-of-mass frame of the pair. The problem is elaborated in detail in \cite{Maj:2009ue}. 

Since we work in the center-of-mass frame of the deuterons, where their motion is assumed to be nonrelativistic, we separate the center-of-mass and relative motion in the nonrelativistic manner. Using the variables
\be 
\label{CM-2}
\left\{ \begin{array}{l}
{\bf R}\equiv \frac{1}{2}({\bf r}_1 + {\bf r}_2 ) ,
\\[1mm]
{\bf r} \equiv  {\bf r}_1-{\bf r}_2 ,
\end{array} \right.
~~~~~~~~~~
\left\{ \begin{array}{l}
{\bf P} \equiv  {\bf p}_1 + {\bf p}_2 ,
\\[1mm]
{\bf q} \equiv  \frac{1}{2}({\bf p}_1 - {\bf p}_2 ) ,
\end{array} \right.
\ee
and writing down the wave function as
\be
\label{wave-fun-DD}
\psi({\bf r}_1, {\bf r}_2) = e^{i{\bf P}{\bf R}} \, \phi_{\bf q} ({\bf r}) ,
\ee
where $\phi_{\bf q} ({\bf r}) $ is the wave function of relative motion of the two deuterons, the correlation function (\ref{fun-corr-D-D-ele}) becomes
\be
\label{fun-corr-D-D-ele-final}
\mathcal{R}({\bf q}) 
= \int d^3 r  \, D_r({\bf r}) \, |\phi_{\bf q}({\bf r})|^2 
\ee
with the `relative' source function given as
\be
\label{relative-source}
 D_r({\bf r}) \equiv \int d^3 R \, 
D ({\bf R} - \!\! \begin{array}{c}\frac{1}{2}\end{array}\!{\bf r}) \, 
D ({\bf R} + \!\! \begin{array}{c}\frac{1}{2}\end{array}\!{\bf r}) .
\ee
If the single-particle source is Gaussian
\be
\label{single-source-Gauss}
D({\bf r}) = \Big(\frac{1}{2 \pi R_s^2} \Big)^{3/2} e^{-\frac{{\bf r}^2}{2R_s^2}} ,
\ee
the root-mean-square (RMS) radius of the source is $\sqrt{3}R_s$, and the relative source equals
\be
\label{relative-source-Gauss}
D_r({\bf r}) = \Big(\frac{1}{4 \pi R_s^2} \Big)^{3/2} e^{-\frac{{\bf r}^2}{4R_s^2}} .
\ee
The relative source is obviously normalized to unity; that is, $\int d^3 r \, D_r({\bf r}) =1$.

\section{Deuterons treated as proton-neutron bound states}
\label{sec-D-bound}

If a deuteron is treated as a neutron-proton bound state created due to final-state interactions similarly to the  $D$-$D$ correlation, the correlation function $\mathcal{R}({\bf p}_1 , {\bf p}_2)$ is defined as
\be
\label{DD-corr-fun-D-bound-def}
\frac{dP_{DD}}{d^3p_1 \, d^3p_2} = 2 \mathcal{R}({\bf p}_1 , {\bf p}_2) \, 
\mathcal{A}^2 \bigg(\frac{dP_p}{d^3p_p} \bigg)^2 \bigg(\frac{dP_n}{d^3p_n} \bigg)^2 ,
\ee
where ${\bf p}_n = {\bf p}_p \approx {\bf p}_1/2 \approx {\bf p}_2/2$. The deuteron formation rate $\mathcal{A}$ is determined by the relation
\be
\label{D-n-p}
\frac{dP_D}{d^3p} = {\cal A} \, \frac{dP_p}{d^3(p/2)}  \frac{dP_n}{d^3(p/2)} .
\ee
The factor of two occurs on the right-hand-side of Eq.~(\ref{DD-corr-fun-D-bound-def}) for the following reason. When we have one neutron-proton pair, the probability to create a deuteron is given by Eq.~(\ref{D-n-p}). When we have two protons $(p_1,p_2)$  and two neutrons $(n_3,n_4)$, the probability to have two deuterons is a product of two expressions (\ref{D-n-p}) multiplied by the factor two because the two deuterons can be built in two ways: $D(p_1,n_3)~\&~D(p_2,n_4)$ and  $D(p_1,n_4)~\&~D(p_2,n_3)$.   

The deuteron formation rate is given as (see {\it e.g.} \cite{Mrowczynski:1992gc})
\be
\label{D-rate-relative}
\mathcal{A} 
= \frac{3}{4} (2 \pi)^3 \int d^3r \, D_r ({\bf r}) |\varphi_D({\bf r})|^2 ,
\ee
where $\varphi_D({\bf r})$ is the wave function of relative motion of the neutron and proton which form the deuteron. The nucleons are assumed to be unpolarized and the spin factor $3/4$ takes into account that there are three spin states of a spin-one deuteron and 4 spin states of a nucleon pair. The factor $(2 \pi )^3$ results from the natural units with $\hbar=1$ we use. If we used $h=1$, the factor would be absent. 

The correlation function of two deuterons with relative momentum ${\bf q}$ is determined by the equation
\ba
\label{fun-corr-D-D-bound}
2\mathcal{R}({\bf q}) \, \mathcal{A}^2 
= \frac{3^2}{4^2} (2 \pi )^6
\sum_{i=1}^2  \int d^3 r_1 \, d^3 r_2 \, d^3 r_3 \, d^3 r_4 \, 
D({\bf r}_1) \, D({\bf r}_2) \, D({\bf r}_3) \, D({\bf r}_4)  |\psi_i({\bf r}_1, {\bf r}_2, {\bf r}_3, {\bf r}_4)|^2 .
\ea
The spin factor $3/4$ in the formula (\ref{fun-corr-D-D-bound}) has the same origin as that in Eq.~(\ref{D-rate-relative}). The protons and neutrons are labeled with the indices 1, 2 and 3, 4, respectively; $\psi_1({\bf r}_1, {\bf r}_2, {\bf r}_3, {\bf r}_4)$ is the wave function of two deuterons with the nucleon content $D(p_1,n_3)~\&~D(p_2,n_4)$ and $\psi_2({\bf r}_1, {\bf r}_2, {\bf r}_3, {\bf r}_4)$ is the wave function of $D(p_1,n_4)~\&~D(p_2,n_3)$. In principle, one should sum up the the functions $\psi_1$ and $\psi_2$ and take the modulus square. Then, the interference term of the wave functions shows up. We have not found a way to compute analytically the term which is expected to be small. So, equation (\ref{fun-corr-D-D-bound}) neglects the term. As we explain below Eq.~(\ref{fun-corr-D-D-bound-final}), the final formula of the correlation function would be very complicated if one took into account the interference term.

We note that our comments about the time dependence of the source function and the reference frame we use, which are formulated in the context of formula (\ref{fun-corr-D-D-ele}), apply equally to the deuteron formation rate (\ref{D-rate-relative}) and equation (\ref{fun-corr-D-D-bound}). In short, the source function effectively takes into account the time duration of the emission processes and we work in the center-of-mass frame of the four-nucleon system where the motion of the nucleons is nonrelativistic. 

The formation time of a deuteron, which is of the order of the inverse deuteron binding energy, is roughly 100 fm/$c$ and it is much bigger than the space-time extension of particles' sources in high-energy collisions. One asks whether the long formation time of a deuteron influences the source radii inferred from the deuteron formation rate or the $D$-$D$ correlation function. The answer is negative -- the formation time plays no role here. The point is that once the neutron-proton pair becomes an isolated system after its emission from the fireball or at the moment of the fireball freeze-out, the temporal evolution of the neutron-proton wave packet does not change the probability that the pair is in the deuteron energy eigenstate. The same holds for the four-nucleon system -- after the fireball freeze-out the probability that the system in a scattering state of two deuterons is fixed. This is evident if one realizes that the femtoscopic formulas of the deuteron formation rate and the $D$-$D$ correlation function are obtained within the quantum-mechanical sudden approximation; see the classical textbook \cite{Schiff-1968}. The  problem is discussed in more detail in the appendix. 

To compute the contribution to the correlation function (\ref{fun-corr-D-D-bound}), which comes from $\psi_1$, we introduce the variables 
\be 
\label{variables-4}
\left\{ \begin{array}{ll}
{\bf R}\equiv \frac{1}{4}({\bf r}_1 + {\bf r}_2 + {\bf r }_3+ {\bf r }_4) ,
\\[1mm]
{\bf r}_{13} \equiv  {\bf r}_1-{\bf r}_3 ,
\\[1mm]
{\bf r}_{24} \equiv  {\bf r}_2-{\bf r}_4 ,
\\[1mm]
{\bf r} \equiv \frac{1}{2}({\bf r}_1 + {\bf r }_3)
- \frac{1}{2}({\bf r}_2 + {\bf r}_4),
\end{array} \right.
~~~~~~~~~~
\left\{ \begin{array}{ll}
{\bf r}_1 = {\bf R}+ \frac{1}{2}({\bf r} + {\bf r }_{13}) ,
\\[1mm]
{\bf r}_2 = {\bf R} -\frac{1}{2}({\bf r} - {\bf r }_{24}) ,
\\[1mm]
{\bf r}_3 ={\bf R}  + \frac{1}{2}({\bf r} - {\bf r }_{13})  ,
\\[1mm]
{\bf r}_4 ={\bf R} - \frac{1}{2}({\bf r} + {\bf r }_{24}) .
\end{array} \right.
\ee
One checks that the Jacobian of the variable transformation equals unity. 

Writing down the wave function as
\be
\label{wave-fun-ppnn}
\psi_1({\bf r}_1, {\bf r}_2, {\bf r}_3, {\bf r}_4) = e^{i{\bf P}{\bf R}} \, \phi_{\bf q}({\bf r}) \, 
 \varphi_D ({\bf r}_{13}) \, \varphi_D ({\bf r}_{24}) ,
\ee
and using the Gaussian source (\ref{single-source-Gauss}), the integral over the center-of-mass position ${\bf R}$  in Eq.~(\ref{fun-corr-D-D-bound}) gives
\be
\label{relative-sources-3}
\int d^3 R \,D ({\bf r}_1) \, D({\bf r}_2) \, D({\bf r}_3)  \, D({\bf r}_4)
=   D_r({\bf r}_{13})  \, D_r({\bf r}_{24})  \, D_{4r}({\bf r}) ,
\ee
where $ D_r({\bf r})$ is again given by Eq.~(\ref{relative-source-Gauss}) and the source function $D_{4r}({\bf r})$ equals
\be
\label{source-D-D}
\mathcal{D}_{4r}({\bf r}) = \Big(\frac{1}{2\pi R_s^2} \Big)^{3/2} e^{-\frac{{\bf r}^2}{2R_s^2}} .
\ee
The function is normalized to unity; that is, $\int d^3 r \, D_{4r}({\bf r}) = 1$.

To compute the contribution to the correlation function (\ref{fun-corr-D-D-bound}), which comes from $\psi_2$, we introduce the variables analogous to (\ref{variables-4}) but ${\bf r}_3 \leftrightarrow {\bf r}_4$. Further calculations are the same. Combining the contributions from $\psi_1$ and $\psi_2$, we find out that due to the integration over ${\bf R}$ in the right-hand-side of Eq.~(\ref{fun-corr-D-D-bound}), the square of the formation rate (\ref{D-rate-relative}) factors out. Consequently, the factor $2{\cal A}^2$, which is also present on the left-hand side of Eq.~(\ref{fun-corr-D-D-bound}), drops out and the correlation function equals
\be
\label{fun-corr-D-D-bound-final}
\mathcal{R}({\bf q}) = \int d^3 r \, D_{4r}({\bf r}) \, |\phi_{\bf q} ({\bf r})|^2.
\ee

The formula (\ref{fun-corr-D-D-bound-final}) has the same form as (\ref{fun-corr-D-D-ele-final}) but the source function differs. When deuterons are directly emitted from the fireball as elementary particles the radius of the deuteron source is the same as the radius of the proton source. When deuterons are formed only after the emission of nucleons, the source becomes effectively bigger because the deuteron formation is a process of spatial extent. More quantitatively, the source radius of deuterons treated as bound states is bigger by the factor $\sqrt{2}\approx 1.41$ than that of `elementary' deuterons. 

We note that if the interference term discussed below Eq.~(\ref{fun-corr-D-D-bound}) is not neglected, the factorization does not occur, and the  $D$-$D$ correlation function is not given by Eq.~(\ref{fun-corr-D-D-bound-final}) but it gets a complicated form. In particular, it depends on the deuteron formation rate ${\cal A}$.

\section{Computation of the correlation function}
\label{sec-corr-fun}

To compute the correlation function given by Eq.~(\ref{fun-corr-D-D-ele-final}) or Eq.~(\ref{fun-corr-D-D-bound-final}) one needs the wave function $\phi_{\bf q}({\bf r})$ of the relative motion of two deuterons.

If the Coulomb interaction is absent but there is a short-range strong interaction, the wave function can be chosen, as proposed in \cite{Lednicky:1981su}, in the asymptotic scattering form
\be 
\label{scatt-wave-fun}
\phi_{\bf q}({\bf r}) = e^{iqz}+f(q)\frac{e^{iqr}}{r} ,
\ee
where $q \equiv |{\bf q}|$ and $f(q)$ is the $s$-wave (isotropic) scattering amplitude chosen as 
\be
\label{amplitude}
f(q)=\frac{-a}{1+iq a} ,
\ee
with $a$ being the scattering length which in general is a complex number. 

\begin{figure}[t]
\begin{minipage}{87mm}
\centering
\includegraphics[scale=0.29]{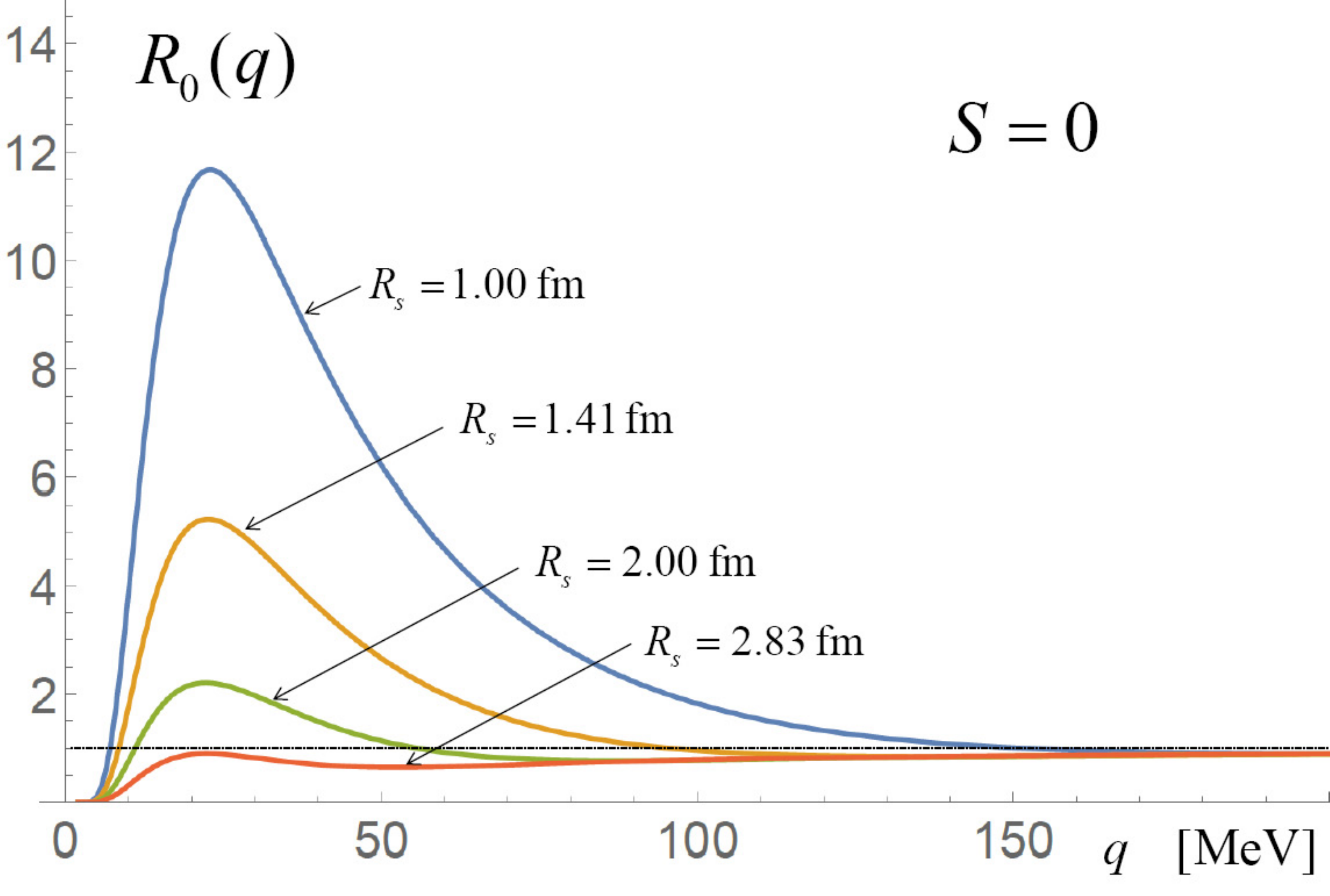}
\vspace{-6mm}
\caption{The  $D$-$D$ correlation function of $S=0$ for four values of $R_s$}
\label{Fig-D-D-S0}
\end{minipage}
\hspace{2mm}
\begin{minipage}{87mm}
\centering
\vspace{9mm}
\includegraphics[scale=0.24]{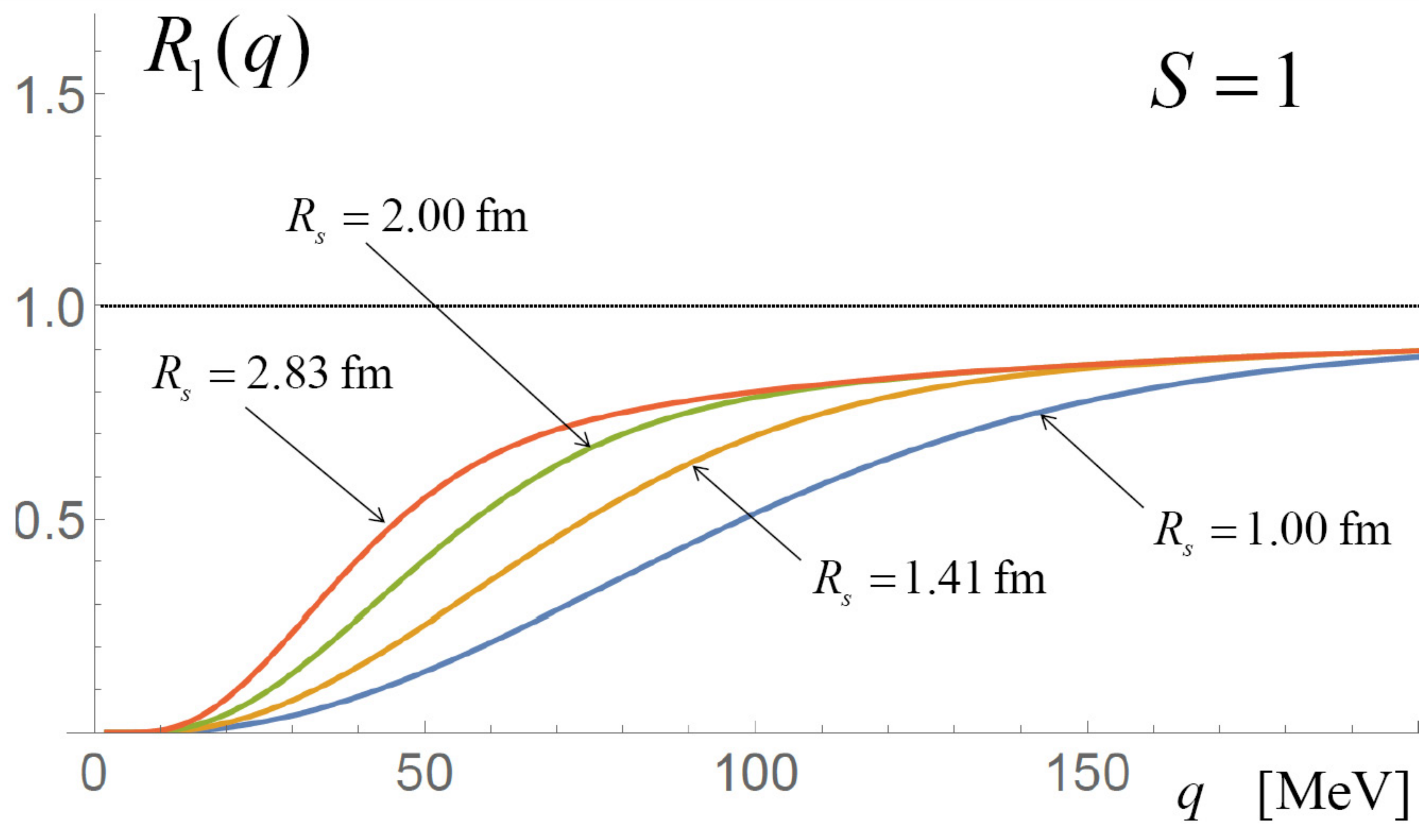}
\vspace{-3mm}
\caption{The $D$-$D$ correlation function of $S=1$ for four values of $R_s$}
\label{Fig-D-D-S1}
\end{minipage}
\hspace{2mm}
\end{figure}

In the case of the $D$-$D$ correlation function we deal with a pair of identical particles and consequently the spatial wave function (\ref{scatt-wave-fun}) has to be symmetrized or antisymmetrized. The total spin $S$ of a $D$-$D$ pair in an $s$-wave can be 0, 1, or 2. The spin wave function of the pair is symmetric with respect to the interchange of the deuterons for $S=0,2$ and antisymmetric for $S=1$. Since the complete wave function must be symmetric with respect to the interchange of the deuterons, the spatial wave function must be symmetric for $S=0,2$ and antisymmetric for $S=1$. Therefore, the wave function (\ref{scatt-wave-fun}) should be replaced by 
\be 
\label{scatt-wave-fun-sym} 
\phi_{\bf q}({\bf r}) \to \frac{1}{\sqrt{2}}\Big(\phi_{\bf q}({\bf r}) + (-1)^S \phi_{\bf q}(-{\bf r}) \Big)
= \frac{1}{\sqrt{2}}
 \bigg(e^{iqz} +(-1)^S e^{-iqz} + \big(1+(-1)^S\big)  f(q)\frac{e^{iqr}}{r} \bigg) ,
\ee
where we have taken into account that the $s$-wave scattering amplitude is symmetric under the mirror reflection ${\bf r} \to -{\bf r}$. Consequently, the effect of interaction shows up for $S=0,2$ but is absent for $S=1$.

Let us compute the correlation function (\ref{fun-corr-D-D-ele-final}) with the Gaussian source (\ref{relative-source-Gauss}).
For $S=1$ the correlation function coincides with that of noninteracting identical fermions and it equals 
\be
\mathcal{R}_1({\bf q}) 
= 1 - e^{-4R_s^2 q^2} .
\ee
For $S=0,~2$ the correlation function (\ref{fun-corr-D-D-ele-final}) is found as
\be
 \label{fun-corr-S=02}
\mathcal{R}_{0,2}({\bf q}) 
= 1+ e^{-4R_s^2 q^2} +  \frac{|f(q)|^2}{R_s^2} - \frac{\Im f(q) }{R_s^2 q} \, (1 - e^{-4R_s^2 q^2}) 
+ \frac{ \Re f(q)}{\pi^{1/2}R_s^3 q}  \int^\infty_0 dr \,e^{-\frac{r^2}{4R_s^2}} \sin(2qr) ,
\ee
where the remaining integral needs to be taken numerically. 

Assuming that deuterons are unpolarized, the  $D$-$D$ correlation function should be averaged over the spin states as
\be
\mathcal{R}({\bf q})  = \frac{1}{9} \mathcal{R}_0({\bf q}) 
+ \frac{3}{9} \mathcal{R}_1({\bf q}) + \frac{5}{9} \mathcal{R}_2({\bf q}) ,
\ee
where $\mathcal{R}_0({\bf q})$, $\mathcal{R}_1({\bf q})$, and $\mathcal{R}_2({\bf q})$ are the correlation functions corresponding to $S=0,~1,~2$, respectively. The weight factors 1/9, 3/9, and 5/9 reflect the numbers of spin states in
the three channels.

Since we deal with charged particles, the formula (\ref{scatt-wave-fun}) should be modified because the long-range electrostatic interaction influences both the incoming and outgoing waves. However, the Coulomb effect can be approximately taken into account \cite{Gmitro:1986ay} by multiplying the correlation function by the Gamow factor that equals
\be 
\label{Gamow}
G(q) = {2 \pi \over a_B q} \,
{1 \over {\rm exp}\big({2 \pi \over a_B q}\big) - 1} ,
\ee
where $a_B$ is the Bohr radius of the deuteron-deuteron pair. Since $a_B^{-1} = \mu \alpha$ with $\mu$ and $\alpha$ being the reduced mass of the $D$-$D$ system and the fine-structure constant, $a_B = 28.8~{\rm fm}$. The Gamow factor is the modulus squared of the exact Coulomb wave function of two charge particles taken at zero distance. The Gamow factor properly takes into account the Coulomb interaction as long as $a_B$ is much bigger that the source radius. We identify the latter quantity with the RMS radius which is $\sqrt{3}R_s$. So, the condition of applicability of the Gamow factor is $R_s \ll  16.6$ fm.

We are not aware of experimentally obtained scattering lengths of the  $D$-$D$ system but there are rather reliable calculations \cite{Filikhin:2000a,Filikhin:2000b}, see also \cite{Carew:2021jai}. We have used the following values of the scattering lengths in the singlet ($S=0$) and quintet ($S=2$) states  
\be
\label{scattering-lengths}
a_0 = (10.2 +0.2i) ~{\rm fm}, ~~~~~~~~~~~~~~~~ a_2 = 7.5 ~{\rm fm} 
\ee
obtained in \cite{Filikhin:2000a,Filikhin:2000b}. Unfortunately, an uncertainty of the theoretical results is not given. The quintet scattering length is also found in \cite{Carew:2021jai} as $7.8 \pm 0.3$~fm. However, the computation scheme is different than in \cite{Filikhin:2000a,Filikhin:2000b} and the scattering length of the $S=0$ channel is not given. Assuming that the uncertainty of the scattering lengths is 0.3 fm the computed correlation functions remain almost unchanged when the scattering lengths are varied within the errors. 

\begin{figure}[t]
\begin{minipage}{87mm}
\centering
\includegraphics[scale=0.24]{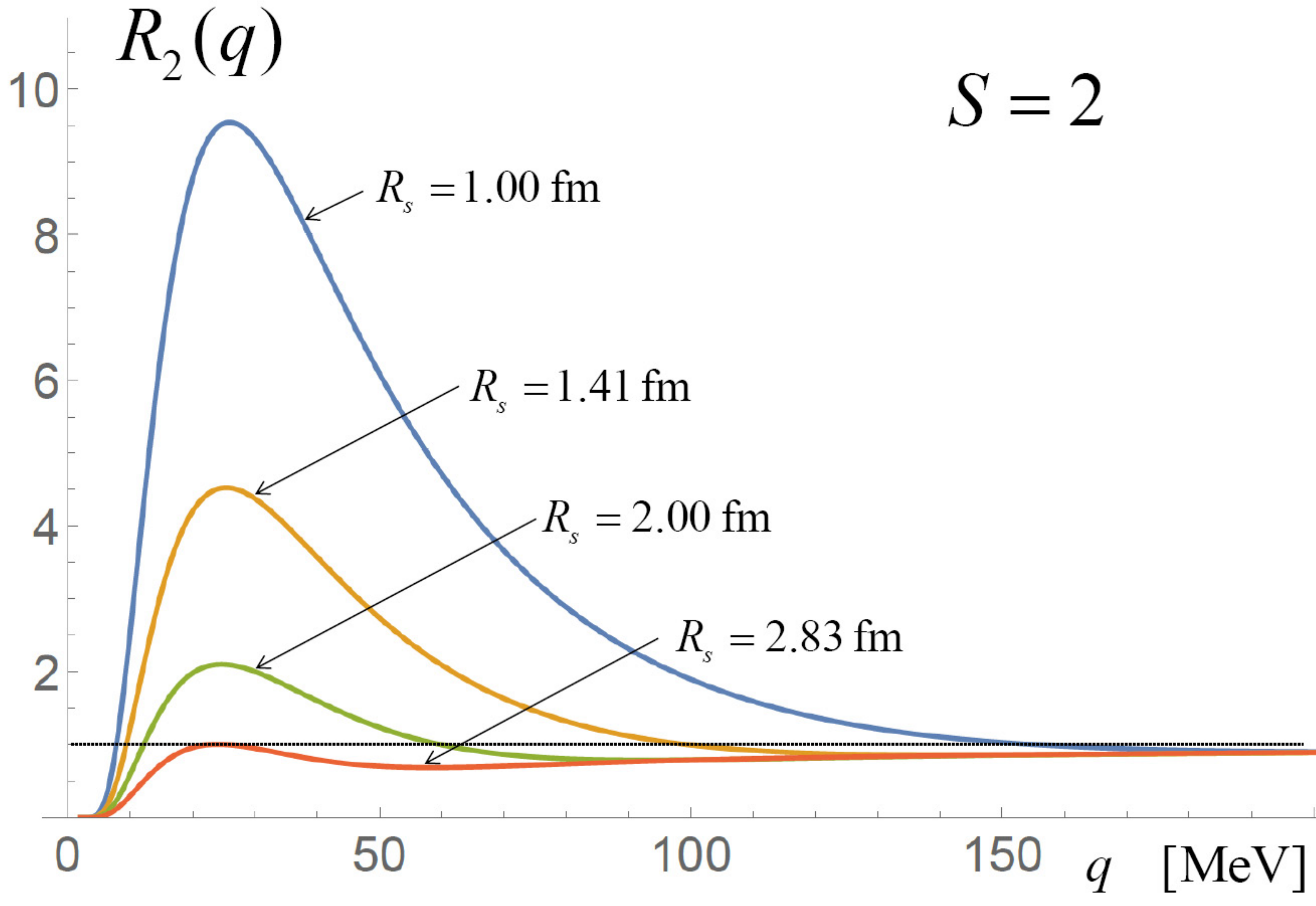}
\vspace{-2mm}
\caption{The  $D$-$D$ correlation function of $S=2$ for four values of $R_s$}
\label{Fig-D-D-S2}
\end{minipage}
\hspace{2mm}
\begin{minipage}{87mm}
\centering
\vspace{3mm}
\includegraphics[scale=0.27]{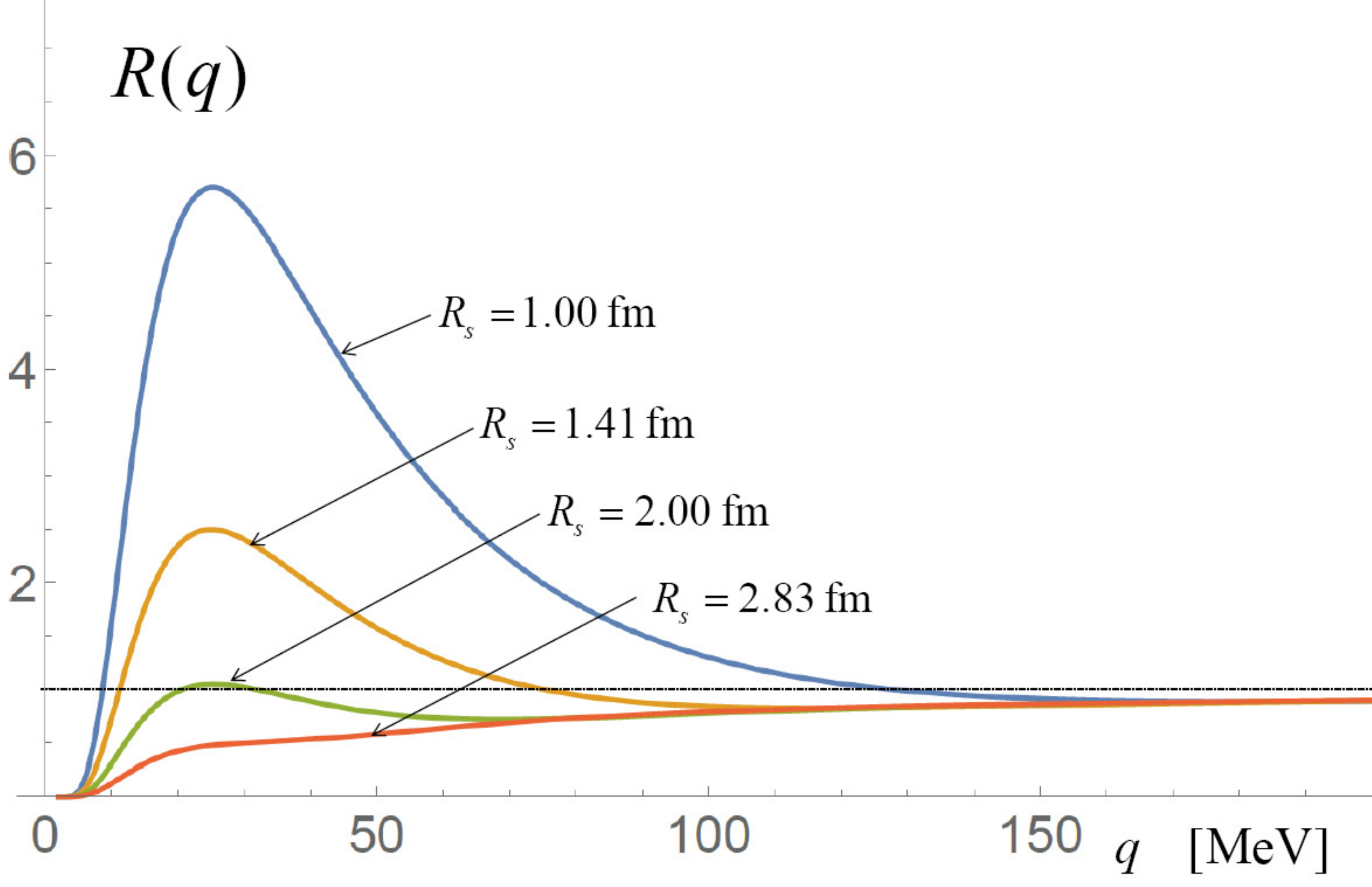}
\vspace{-3mm}
\caption{The spin-averaged $D$-$D$ correlation function for four values of $R_s$}
\label{Fig-D-D-ave}
\end{minipage}
\hspace{2mm}
\end{figure}

Using the numbers (\ref{scattering-lengths}) and the deuteron mass $m_D = 1876$ MeV, we have calculated according to Eq.~(\ref{fun-corr-D-D-ele-final}) the correlation functions $\mathcal{R}_0(q)$, $\mathcal{R}_1(q)$, $\mathcal{R}_2(q)$ and the spin averaged $\mathcal{R}(q)$ for $R_s=1.00, ~1.41,~2.00$ and 2.83 fm. The results are shown in Figs.~\ref{Fig-D-D-S0}, \ref{Fig-D-D-S1}, \ref{Fig-D-D-S2} and \ref{Fig-D-D-ave}, respectively. The correlation functions presented in the figures take into account the strong interaction and Coulomb repulsion together with the effect of indistinguishability of deuterons. 

The source radii are chosen in such a way that $R_s = 2.83 = \sqrt{2} \cdot 2 = (\sqrt{2})^2 \cdot 1.41 =  (\sqrt{2})^3 \cdot 1.00$ fm. Therefore, the scenarios of deuterons directly emitted from the fireball and of deuterons formed due to final-state interactions correspond to each pair of neighboring curves in Figs.~\ref{Fig-D-D-S0}, \ref{Fig-D-D-S1}, \ref{Fig-D-D-S2} and \ref{Fig-D-D-ave}. One has to experimentally distinguish the two curves to distinguish the two mechanisms of deuteron production.  

As mentioned in the introduction, the  $D$-$D$ correlation function was measured in nucleus-nucleus collisions at the energy of a few tens MeV per nucleon \cite{Chitwood:1985zz,Pochodzalla:1987zz,Gourio:2000tn} and in proton-nucleus collisions at 500 MeV \cite{Cebra:1989xe}, see also the review \cite{Verde:2006dh} and references therein. The $D$-$D$ correlation functions we computed approximately agree with those measured in the low-energy nucleus-nucleus collisions. However, the comparison is not critical. The point is that the source radii obtained in these measurements are rather large, corresponding to our $R_s$  from the interval 3-7 fm. Then, the correlation function is dominated by the Coulomb repulsion which is taken into account in our calculations by means of the Gamow factor. As explained below Eq.~(\ref{Gamow}), the method is accurate for $R_s \ll  16.6$ fm. So, our calculations are not very precise for the radii which satisfy the condition with a small margin. However, the correlation functions measured for big sources are not relevant for our proposal; in such a case the functions weakly depend of the source radius $R_s$ and it is difficult to distinguish $R_s$ from $\sqrt{2} R_s$. 

Let us also that in agreement with our expectations it was found in \cite{Chitwood:1985zz} that the source radius inferred from the $p$-$p$-correlation function is smaller by factor 1.5-2.0 than that from the $D$-$D$ correlation function. However, in the low-energy collisions there are several factors which can contribute to the source size difference. 

We are particularly interested in deuterons produced at midrapidity at the LHC because their origin is under debate. The radius of the proton source measured at the LHC varies from $R_s \approx 1.1~{\rm fm}$ in $p$-$p$ collisions \cite{Acharya:2020dfb,Acharya:2018gyz} to $R_s \approx 4.0~{\rm fm}$ in central Pb-Pb collisions \cite{Adam:2015vja}. So, we are in the domain of source sizes where, as Fig.~\ref{Fig-D-D-ave} shows, the $D$-$D$ correlation function is sensitive to the source radius. Consequently, it can be precisely measured. Then, a systematic measurement of $p$-$p$,  $p$-$D$, and $D$-$D$ correlation functions can tell us whether the deuterons are directly emitted from the fireball or are formed due to final-state interactions. The $p$-$p$ correlation function has been already measured with a high accuracy \cite{Acharya:2018gyz,Acharya:2020dfb,Adam:2015vja}, and a measurement of the $p$-$D$ function is feasible \cite{Fabbietti:2020}, but it is a real challenge to obtain the $D$-$D$ correlation function. Since a production of $^4{\rm He}$ is observed at midrapidity in Pb-Pb collisions at the LHC \cite{Acharya:2017bso}, $D$-$D$  pairs certainly occur. However, it will be difficult to collect a sufficient statistics of the pairs to obtain the $D$-$D$ correlation function.

Our findings remain relevant for relativistic nucleus-nucleus collisions at the energies of a few GeV in the center-of-mass of nucleon-nucleon pairs. Then, nucleons and light nuclei are abundantly produced at midrapidity but the excitation energy per baryon of the fireball significantly exceeds the nuclear binding energy. Since the direct emission of deuterons from the fireball is suppressed if not excluded at all, we expect that the source radius inferred from the $D$-$D$ correlation function is bigger by the factor of $\sqrt{2}$ than that obtained from the $p$-$p$ correlation function. By performing the measurements for noncentral collisions we can deal with relatively small particle sources when the $D$-$D$ correlation function strongly depends on the source radius.

It would also be interesting to use our formulas in a situation where deuterons, as protons, are expected to be directly emitted from a source. Such a situation presumably occurs in the fragmentation domain of relativistic heavy-ion collisions, that is, for deuterons with rapidities close to that of the projectile or target. Then, the deuterons come from the spectator parts of colliding nuclei which are much less excited than the participant parts. The proposed measurement is difficult, if possible at all, in collider experiments but it is certainly feasible in fixed-target experiments.

\section{Summary and Conclusions}
\label{sec-conclusions}

We have derived a formula of the $D$-$D$ correlation function in two ways. At the beginning the deuterons have been treated as elementary particles and further on as neutron-proton bound states. In the first case the deuterons are directly emitted from the source and in the second case the deuterons are formed only after the nucleons are emitted from the source. Then, the deuteron formation is simultaneous with a generation of $D$-$D$ correlations. We have found that the source radius of deuterons formed due to final-state interactions is bigger by the factor of $\sqrt{2}$ than that of directly emitted deuterons. Previously we found the analogous result for the $p$-$D$ correlation function but the effect of enlargement of the source radius, which happens due to the space-time extension of the deuteron formation process, was by the smaller factor of $\sqrt{4/3}\,$ \cite{Mrowczynski:2019yrr}. 

To check whether the enlargement of the source radius is a measurable effect we have computed the $D$-$D$ correlation function. The Bose-Einstein statistics of deuterons and the $s$-wave scattering due the strong interaction and Coulomb repulsion have been taken into account. The correlation function is sensitive to the source radius for sufficiently small sources with the RMS radii smaller than 3.5 fm. Otherwise the correlation function is dominated by the Coulomb repulsion and it weakly depends on the source radius. 

We have discussed our findings in the context of existing data from low-energy nuclear collisions and have also considered measurements which can clarify a mechanism of deuteron production making use of our findings.

\section*{Acknowledgments}

This work was partially supported by the National Science Centre, Poland, under grant 2018/29/B/ST2/00646.

\appendix*

\section{Deuteron Formation Time}

The formation time of a deuteron, which is of order 100 fm/$c$, plays an important role when multiple interactions of a deuteron are considered. The two interaction processes cannot be treated as independent from each other if they are not separated by a time interval which is much longer than the formation time. We explain here why the formation time does not influence the source radii inferred from the deuteron formation rate or the $D$-$D$ correlation function. 

Let us first discuss the case of a single deuteron. One could argue that the deuteron formation rate is determined by the relative distance of the neutron and proton not at the moment of the fireball freeze-out, as given by Eq.~(\ref{D-rate-relative}), but at a later time which includes the deuteron formation time. Since the formation time is much longer than the space-time radius of the source, the effect could be sizable. However, this is not the case as follows from the derivation of the deuteron formation rate formula (\ref{D-rate-relative}), which relies on the quantum-mechanical sudden approximation, see the classical textbook \cite{Schiff-1968}.

A rather detailed derivation of the deuteron formation rate is given in the old paper by one of us \cite{Mrowczynski:1987oid}. Here we discuss only the sudden approximation aspect of the derivation. The fireball freezes out or decays at the time $t_f$, which is identified with a sudden change of the system's Hamiltonian. For $t < t_f$ a neutron-proton pair interacts with fireball constituents and for $t > t_f$ the pair is an isolated system. According to the sudden approximation, the probability that the neutron-proton pair, which is described by the wave function $\psi (t,{\bf r})$, is in a deuteron energy eigenstate $\psi_D (t,{\bf r})$ equals
\be
P = \Big| \int d^3 r \, \psi^* (t_f,{\bf r}) \, \phi_D ({\bf r}) \Big|^2  = 
\int d^3 r \, d^3 r' \, \psi^* (t_f,{\bf r}) \, \psi (t_f,{\bf r}') \, \psi_D (t_f,{\bf r}) \, \psi_D^* (t_f,{\bf r}') .
\ee
To simplify the discussion we consider here only the relative motion of the neutron and proton, and consequently the wave functions $\psi (t,{\bf r})$ and $\psi_D (t,{\bf r})$ describe only the relative motion. 

Since $\psi_D (t,{\bf r})$ is the energy eigenstate of the eigenvalue $E$ it is of the from 
\be
\psi_D (t,{\bf r}) = e^{-iEt} \phi_D ({\bf r}) ,
\ee
which when substituted in the probability expression gives
\be
P = 
\int d^3 r \, d^3 r' \, \psi^* (t_f,{\bf r}) \, \psi (t_f,{\bf r}') \, \phi_D ({\bf r}) \, \phi_D^* ({\bf r}') .
\ee
As one observes, the time dependence of the deuteron wave function disappears.

Since the neutron-proton pair in a fireball is a part of a bigger system it should be described not by the wave function $\psi (t,{\bf r})$ but rather by the density matrix $\rho(t,{\bf r},{\bf r}')$. So, we replace $\psi^* (t_f,{\bf r}) \, \psi (t_f,{\bf r}')$  by $\rho(t_f,{\bf r},{\bf r}')$ and the probability becomes 
\be
P = \int d^3 r \, d^3 r' \, \rho(t_f,{\bf r},{\bf r}') \, \phi_D ({\bf r}) \, \phi_D^* ({\bf r}') .
\ee
As explained in \cite{Mrowczynski:1987oid}, the assumption, which leads to the final deuteron formation rate formula (\ref{D-rate-relative}), is equivalent to the assumption that the density matrix is diagonal that is 
\be
\rho(t_f,{\bf r},{\bf r}')  \sim \delta^{(3)}({\bf r}-{\bf r}')  D({\bf r}) ,
\ee
where $D({\bf r})$ is the source function being the probability distribution of neutron-proton relative distance at freeze-out. Thus, one finds
\be
P \sim \int d^3 r \, D({\bf r}) \big| \phi_D ({\bf r}) \big|^2 ,
\ee
which is up the spin and normalization factors our Eq.~(11). 

The neutron-proton wave function $\psi (t,{\bf r})$ obviously evolves beyond the freeze-out time $t_f$. However, the probability that the pair is in a deuteron energy eigenstate is determined at the freeze-out when the system's Hamiltonian suddenly changes and the pair becomes isolated. Further evolution of the wave function $\psi (t,{\bf r})$ does not change the probability because the function can be expressed as a superposition of the energy eigenfunctions with time independent coefficients. Consequently, the projection of the wave packet $\psi (t,{\bf r})$ on the deuteron energy eigenfunction equals the corresponding coefficient. 

We note that an accuracy of the sudden approximation, which is discussed in detail in the textbook \cite{Schiff-1968}, depends on how fast the system's Hamiltonian changes but not on the time scale of the evolution of $\psi (t,{\bf r})$. 

When one deals not with one but with two deuterons in a scattering state the situation is very similar. The femtoscopic formula of the $D$-$D$ correlation function is obtained by projecting the four-nucleon state at the moment of freeze-out on the wave function of two deuterons in a scattering state. The wave-packet of four nucleons evolves after the freeze-out time but once the four-nucleon system becomes isolated the probability that the four nucleons are in the energy eigenstate of two deuterons does not change in time. Consequently, the deuteron formation time does not influence the source radius inferred from the femtoscopic correlation function.



\begin{thebibliography}{99}

\bibitem{Alper:1973my}
B.~Alper \textit{et al.}
Phys. Lett. B \textbf{46}, 265 (1973).

\bibitem{Henning:1977mt}
S.~Henning \textit{et al.} [British-Scandinavian-MIT Collaboration],
Lett. Nuovo Cim. \textbf{21}, 189 (1978).

\bibitem{Acharya:2019rgc}
S.~Acharya \textit{et al.} [ALICE Collaboration],
Phys. Lett. B \textbf{794}, 50 (2019).

\bibitem{Acharya:2020sfy}
S.~Acharya \textit{et al.} [ALICE Collaboration],
Eur. Phys. J. C \textbf{80}, 889 (2020).

\bibitem{Butler:1963pp} 
S.~T.~Butler and C.~A.~Pearson,
Phys.\ Rev.\  {\bf 129}, 836 (1963).

\bibitem{Schwarzschild:1963zz} 
A.~Schwarzschild and C.~Zupancic,
Phys.\ Rev.\  {\bf 129}, 854 (1963).
 
\bibitem{Mrowczynski:1987oid}
St.~Mr\'owczy\'nski,
J. Phys. G \textbf{13}, 1089 (1987).

\bibitem{Bellini:2020cbj}
F.~Bellini, K.~Blum, A.~P.~Kalweit and M.~Puccio,
Phys. Rev. C \textbf{103}, 014907 (2021).

\bibitem{Andronic:2010qu}
A.~Andronic, P.~Braun-Munzinger, J.~Stachel and H.~Stocker,
Phys.\ Lett.\ B {\bf 697}, 203 (2011).

\bibitem{Cleymans:2011pe}
J.~Cleymans \textit{et al.}
Phys.\ Rev.\ C {\bf 84}, 054916 (2011).

\bibitem{Andronic:2017pug} 
A.~Andronic, P.~Braun-Munzinger, K.~Redlich and J.~Stachel,
Nature {\bf 561}, 321 (2018).

\bibitem{Mrowczynski:2020ugu}
St.~Mr\'owczy\'nski,
Eur. Phys. J. ST \textbf{229}, 3559 (2020).

\bibitem{Lisa:2005dd} 
M.~A.~Lisa, S.~Pratt, R.~Soltz and U.~Wiedemann,
Ann.\ Rev.\ Nucl.\ Part.\ Sci.\  {\bf 55}, 357 (2005).

\bibitem{Mrowczynski:2019yrr}
St.~Mr\'owczy\'nski and P.~S\l o\'n,
Acta Phys. Polon. B \textbf{51}, 1739 (2020).

\bibitem{Chitwood:1985zz}
C.~B.~Chitwood \textit{et al.}
Phys. Rev. Lett. \textbf{54}, 302 (1985).

\bibitem{Pochodzalla:1987zz} 
J.~Pochodzalla {\it et al.},
Phys.\ Rev.\ C {\bf 35}, 1695 (1987).

\bibitem{Gourio:2000tn}
D.~Gourio \textit{et al.} [INDRA Collaboration],
Eur. Phys. J. A \textbf{7}, 245 (2000).

\bibitem{Cebra:1989xe}
D.~A.~Cebra \textit{et al.}
Phys. Lett. B \textbf{227}, 336 (1989).

\bibitem{Verde:2006dh}
G.~Verde, A.~Chbihi, R.~Ghetti and J.~Helgesson,
Eur. Phys. J. A \textbf{30}, 81 (2006).

\bibitem{Koonin:1977fh} 
S.~E.~Koonin,
Phys.\ Lett.\  {\bf 70B}, 43 (1977).

\bibitem{Maj:2009ue} 
R.~Maj and St.~Mr\'owczy\'nski,
Phys.\ Rev.\ C {\bf 80}, 034907 (2009).

\bibitem{Schiff-1968}
L.~I.~Schiff, {\it Quantum Mechanics} (McGraw-Hill, 3rd ed., New Your, 1965) \S 35, p. 292. 

\bibitem{Adam:2015vja} 
J.~Adam {\it et al.} [ALICE Collaboration],
Phys.\ Rev.\ C {\bf 92}, 054908 (2015).

\bibitem{Acharya:2020dfb}
S.~Acharya \textit{et al.} [ALICE],
Phys. Lett. B \textbf{811}, 135849 (2020).

\bibitem{Mrowczynski:1992gc} 
St.~Mr\'owczynski,
Phys.\ Lett.\ B {\bf 277}, 43 (1992).

\bibitem{Lednicky:1981su} 
R.~Lednicky and V.~L.~Lyuboshits,
Sov.\ J.\ Nucl.\ Phys.\  {\bf 35}, 770 (1982)  [Yad.\ Fiz.\  {\bf 35}, 1316 (1981)].

\bibitem{Gmitro:1986ay} 
M.~Gmitro, J.~Kvasil, R.~Lednicky and V.~L.~Lyuboshitz,
Czech.\ J.\ Phys.\ B {\bf 36}, 1281 (1986).

\bibitem{Filikhin:2000a}
I.N.~Filikhin and S.L.~Yakovlev,
Phys. Atom. Nucl. {\bf 63}, 55 (2000) [Yad. Fiz. {\bf 63}, 63 (2000)].

\bibitem{Filikhin:2000b}
I.N.~Filikhin and S.L.~Yakovlev,
Phys. Atom. Nucl. {\bf 63}, 216 (2000) [Yad. Fiz. {\bf 63}, 271 (2000)].

\bibitem{Carew:2021jai}
J.~F.~Carew,
Phys. Rev. C \textbf{103}, 014002 (2021).

\bibitem{Acharya:2018gyz} 
S.~Acharya {\it et al.} [ALICE Collaboration],
Phys.\ Rev.\ C {\bf 99}, 024001 (2019).

\bibitem{Fabbietti:2020}
Laura Fabbietti, private communication.

\bibitem{Acharya:2017bso} 
S.~Acharya {\it et al.} [ALICE Collaboration],
Nucl.\ Phys.\ A {\bf 971}, 1 (2018).

\end{thebibliography}
\end{document}